\documentclass[12pt,a4paper]{article}
\usepackage{amsfonts}
\usepackage{amsmath}

\def\beq{\begin{equation}}
\def\eeq{\end{equation}}

\def\dg{\dagger}

\def\bt{\begin{tabular}}
\def\et{\end{tabular}}

\input{tcilatex}

\begin{document}

\author{O. Cherbal$^{(1)}$, M. Drir$^{(1)}$, M. Maamache $^{(2)}$ and D.A.
Trifonov$^{(3)}$ \\
$^{(1)}${\normalsize \thinspace Faculty of Physics, Theoretical Physics
Laboratory, University of \ \ }\\
{\normalsize Bab-Ezzouar, USTHB, B.P. 32, El Alia, Algiers 16111, Algeria}\\
$^{(2)}$\,{\normalsize Laboratoire de Physique Quantique et Syst\`{e}mes
Dynamiques, }\\
{\normalsize Department of Physics, Setif University, Setif 19000, Algeria.}%
\\
$^{(3)}$\,{\normalsize Institute of Nuclear Research, 72 Tzarigradsko
chauss\'ee,} \\
{\normalsize 1784 Sofia, Bulgaria.}}
\title{Invariants and Coherent States for Nonstationary Fermionic Forced
Oscillator}
\maketitle

\vspace{-99mm}
\noindent e-print arXiv: quant-ph/0903.3312
\vspace{99mm}

\begin{abstract}
The most general form of Hamiltonian that preserves fermionic coherent
states stable in time is found in the form of nonstationary fermion
oscillator. Invariant creation and annihilation operators and related Fock
states and coherent states are built up for the more general system of
nonstationary forced fermion oscillator.

\end{abstract}

\medskip

\section{Introduction}

\ \ \ \ The time evolution of coherent states (CS) has attracted a great
deal of attention since the introduction of Glauber's CS of the harmonic
oscillator \cite{Glauber63a, Glauber63b,Glauber63c}. Of particular interest
has been the determination of the Hamiltonian operator for which an initial
coherent state remains coherent under time evolution. It is established that
this Hamiltonian has the form of the nonstationary bosonic forced oscillator
Hamiltonian {\normalsize \cite{Glauber66, Mehta66, Mehta67, Stoler75, Kano76}%
: \ \ \ \ \ \ \ } 
\begin{equation}
H_{\mathrm{cs}} = \omega (t)a^{\dg}a+f(t)a^{\dg}+f^{\ast}(t)a+\beta(t),
\label{Hcs}
\end{equation}%
where $\omega (t)$ and $\beta (t)$ are arbitrary real functions of time $t$,
and $f(t)$ is arbitrary complex function. {\normalsize \ }

Our purpose in the present article is to study the dynamical invariants and
time evolution of CS for general (one mode) fermionic Hamiltonian and to
establish the most general form of Hamiltonian which preserves the fermionic
CS under the time evolution.

The organization of the article is as follows. We start with a review in
Sec. II of some main results of time evolution of bosonic forced harmonic
oscillator. In Sec. III we study the temporal stability of fermionic CS and
we show, by using the fermionic analog of the invariant boson ladder
operator method {\cite{MMT70, Holz70, Trif75},} that the most general form
of Hamiltonian that preserves fermionic CS stable in time is in the form of
nonstationary fermion oscillator. In Sec. IV we treat the more general
system of nonstationary forced fermion oscillator (FFO), which is shown to
be the most general one mode fermionic Hamiltonian system. Following the
scheme related to the boson system \cite{MMT70} we construct the dynamically
invariant fermion ladder operators and related Lewis-Riensenfeld Hermitian
invariant \cite{Lewis69}. Using these invariants, we construct fermionic
Fock states and CS for FFO system, which can represent (under appropriate
initial conditions) the exact time-evolution of initial canonical CS.
Finally the relation of the invariant ladder operators method \cite{MMT70,
Holz70} to the Lewis-Riesenfeld method \cite{Lewis69} is briefly described
on the example of FFO. The paper ends with concluding remarks.\ 

\medskip

\section{Canonical CS and their temporal stability}

The standard boson coherent states (CS) (called also Glauber CS, or
canonical CS) are defined as the right eigenstates of the boson (photon)
annihilation operator $a$ {\normalsize \cite%
{Glauber63a,Glauber63b,Glauber63c}}%
\begin{equation*}
a|z\rangle =z|z\rangle
\end{equation*}%
the eigenvalue $z$ being a complex number. The annihilation and creation
operators $a$ and $a^{\dg}$ satisfy the commutation relations$[ a,a^{\dg}] =
aa^{\dg}-a^{\dg}a = 1$. The normalized CS $|z\rangle $ can be constructed in the
from of displaced ground state $|0\rangle$ \cite%
{Glauber63a,Glauber63b,Glauber63c}, 
\begin{equation}
\left\vert z\right\rangle =D(z)\left\vert 0\right\rangle ,\text{ \ }%
D(z)=e^{\left( za^{\dg}-z^{\ast }a\right) },
\end{equation}%
and their expansion in terms of the number states $\left\vert n\right\rangle 
$ reads%
\begin{equation}
\left\vert z\right\rangle =e^{-\frac{\left\vert z\right\vert ^{2}}{2}}%
\overset{\infty }{\underset{n=0}{\tsum }}\frac{z^{n}}{\sqrt{n!}}\left\vert
n\right\rangle
\end{equation}%
\ \ \ \ \ \ \ \ \ 

The problem of temporal stability of bosonic CS is solved by Glauber 
{\normalsize \cite{Glauber66}} and Mehta and Sudarshan {\normalsize \cite%
{Mehta66} }(in the case of one mode CS, and for $n$-mode CS - by Mehta et
al. {\normalsize \cite{Mehta67}}). The result is that the most general
Hamiltonian that preserves an initial CS $\left\vert z\right\rangle $ stable
in later time is of the form of the nonstationary forced oscillator
Hamiltonian $H_{\mathrm{cs}}$, eq. (\ref{Hcs}). The Hamiltonian (\ref{Hcs})
that preserves CS stable is shortly called \textit{coherence Hamiltonian}.
Thus the boson coherence Hamiltonian takes the form of a \textit{%
non-stationary forced oscillator} Hamiltonian. Here "stable" means that the
time evolved state $\left\vert z;t\right\rangle $, 
\begin{equation}
i\frac{d}{dt}\left\vert z;t\right\rangle =H_{cs}\left\vert z;t\right\rangle .
\end{equation}%
remains eigenstate of $a$ possibly with a time-dependent eigenvalue $z(t)$, 
\begin{equation}
a\left\vert z;t\right\rangle =z(t)\left\vert z;t\right\rangle
\end{equation}%
From the latter equation one deduces that, up to a time-dependent phase
factor $exp(i\varphi (t))$, the time-evolved CS $\left\vert z;t\right\rangle 
$ depends on time $t$ through $z(t)$, that is 
\begin{equation}
\left\vert z;t\right\rangle =e^{i\varphi (t)}\left\vert z(t)\right\rangle ,%
\text{ \ \ \ }\left\vert z(t)\right\rangle =e^{a^{\dg}z(t)-z^{\ast
}(t)a}\left\vert 0\right\rangle
\end{equation}%
One says that for system with Hamiltonian ({\normalsize \ref{Hcs}}) an
initial CS remains CS all the later time {\normalsize \cite{Glauber66,
Mehta66} }(or remains \textit{temporally stable}). For the Hamiltonian
system ({\normalsize \ref{Hcs}}) the time dependent eigenvalue value $z(t)$
obeys the equation {\normalsize \cite{Glauber66, Mehta66}} 
\begin{equation*}
i\dot{z}=\omega (t)z+f(t)
\end{equation*}%
the solution of which takes the explicit form\ ($z=z(0)$) 
\begin{eqnarray}
z(t)\ &=& \tilde{\beta}(t)z+\tilde{\gamma}(t),\text{ }\tilde{\beta}%
(t)=e^{-i\tint_{0}^{t}\omega (t^{^{\prime }})dt^{^{\prime }}},  \label{8} \\
\tilde{\gamma}(t) &=&-i\left( \tint_{0}^{t}e^{i\tint_{0}^{t^{^{\prime
}}}\omega (\tau )d\tau }f(t^{^{\prime }})dt^{^{\prime }}\right)
e^{-i\tint_{0}^{t}\omega (t^{^{\prime }})dt^{^{\prime }}}
\end{eqnarray}%
In the particular case of constant $\omega $ we have 
\begin{equation*}
z(t)\ = e^{-i\omega _{0}t}\left( z-i\tint_{0}^{t}e^{i\omega _{0}t^{^{\prime
}}}F(t^{^{\prime }})dt^{^{\prime }}\right) .\text{\ }
\end{equation*}%
The forced oscillator system ({\normalsize \ref{Hcs}}) admits linear in
terms of $a$ and $a^{\dg}$ invariant boson annihilation operator $A_{c}(t),$ $%
[A(t),A^{\dg}(t)]=1,$ 
\begin{equation}  \label{010}
A(t) = U(t)aU^{\dg}(t)=\beta (t)a+\gamma (t)\equiv A_{c}, 
\end{equation}%
where $U(t)$ is the unitary evolution operator, and 
\begin{equation*}
\beta (t) = e^{i\tint_{0}^{t}\omega (t^{^{\prime }})dt^{^{\prime }}}=\tilde{%
\beta}^{-1}(t),\text{ \ \ }\gamma (t) = i\tint_{0}^{t}f(t^{^{\prime
}})e^{i\tint_{0}^{t^{^{\prime }}}\omega (\tau )d\tau }dt^{^{\prime }}=-%
\tilde{\gamma}(t).
\end{equation*}%
For any system the time-evolved CS $\left\vert z;t\right\rangle $ are
eigenstates of the corresponding invariant annihilation operator $A(t)$ with
constant eigenvalues $z$, $A(t)$ $\left\vert z;t\right\rangle =z\left\vert
z;t\right\rangle $, and can be represented in the form of invariantly
displaced time-evolved ground state $\left\vert 0;t\right\rangle
=U(t)\left\vert 0\right\rangle ,$ 
\begin{equation}  \label{11}
\left\vert z;t\right\rangle = D(z,A(t))\left\vert 0;t\right\rangle ,\text{ \
\ }D(z,A(t))=e^{A^{\dg}(t)z-z^{\ast }A(t)}. 
\end{equation}%
If $A(t)$ is invariant then $A^{\dg}(t)$ also is, and any other combination of
them is also invariant. In particular $A^{\dg}(t)A(t)$ and $D(z,A(t))$ are
also invariant operators of the forced oscillator ({\normalsize \ref{Hcs}}).
Invariant operators are very useful, since they transform solutions into
solutions, as demonstrated in ({\normalsize \ref{11}}).

The invariant boson ladder operator ({\normalsize \ref{010}}) is a simple
particular case of linear invariants of general quadratic quantum system,
constructed first in {\normalsize \cite{MMT70, Holz70}}. For the
nonstationary quantum oscillator Hermitian quadratic in $a$ and $a^{\dg}$
invariant was constructed and studied by Lewis and Riesenfeld {\normalsize 
\cite{Lewis69}}. Using these properties of the invariants it was shown 
{\normalsize \cite{Trif75}} that a given Hamiltonian $H$ preserves the
temporal stability of CS $\left\vert z\right\rangle $ if and only if it
admits invariant of the form $A_{c}=\beta (t)a+\gamma (t)$. The general form
of such Hamiltonian coincides with Glauber-Mehta-Sudarshan coherence
Hamiltonian ({\normalsize \ref{Hcs}}).

\section{Temporal stability of canonical fermion CS}

Fermion coherent states (CS) are defined (see {\normalsize \cite{Abe89,
Maam92, Junker98, Cahill99}}) as eigenstates of the fermion annihilation
operator $b$, 
\begin{equation}
b\left\vert \zeta \right\rangle =\zeta \left\vert \zeta \right\rangle ,\text{%
{}}  \label{b|z>}
\end{equation}%
where the eigenvalue $\zeta $ is a Grassmannian variable: $\zeta ^{2}=0$, $\
\zeta \zeta ^{\ast }+\zeta ^{\ast }\zeta =0$. Recall the fermion algebra: 
\begin{equation}
\left\{ b,b^{\dg}\right\} \equiv bb^{\dg}+b^{\dg}b=1,\text{ \ }b^{2}=b^{\dg}{}^{2}=0.
\label{013}
\end{equation}%
For definiteness eigenstates of fermion ladder operator $b$ should be called 
\textit{canonical fermion CS}. This is in analogy to the eigenstates of
boson annihilaion operator $a$, which are known as Glauber CS and \textit{%
canonical boson CS} as well. In terms of the Grasmann eigenvalues $\zeta $
many of the properties of $\left\vert \zeta \right\rangle $ repeat the
corresponding ones of the bosonic CS $\left\vert z\right\rangle $ 
{\normalsize \cite{Cahill99}}. In particular one has 
\begin{equation}
\left\vert \zeta \right\rangle =D(\zeta )\left\vert 0\right\rangle =e^{-%
\frac{1}{2}\zeta ^{\ast }\zeta }\left( \left\vert 0\right\rangle -\text{ }%
\zeta \left\vert 1\right\rangle \right) \,.  \label{|z>}
\end{equation}%
\begin{equation}
\int d\zeta ^{\ast }d\zeta \,|\zeta \rangle \langle \zeta |\text{ }=1,\text{
\ \ }  \label{compl1}
\end{equation}%
where $D(\zeta )=\exp (b^{\dg}\zeta -\zeta ^{\ast }b)$, $\left\vert
0\right\rangle $ is the fermionic vacuum, $b\left\vert 0\right\rangle =0$,
and $\left\vert 1\right\rangle $ is the one-fermion state, $\left\vert
1\right\rangle =b^{\dg}\left\vert 0\right\rangle $. The integrations over $%
\zeta $ and $\zeta ^{\ast }$ are performed according to the Berezin rules 
{\normalsize \cite{Cahill99}} 
\begin{equation}
\int d\zeta ^{\ast }d\zeta \zeta \zeta ^{\ast }=1,\int d\zeta ^{\ast }d\zeta
\zeta =\int d\zeta ^{\ast }d\zeta \zeta ^{\ast }=\int d\zeta ^{\ast }d\zeta
1=0.  \label{z ints}
\end{equation}

The temporal stability of the canonical fermion CS is defined in analogy to
the temporal stability of canonical boson CS, namely the evolution of an
initial $|\zeta \rangle $ is stable if the time-evolved state $|\zeta
;t\rangle =U(t)|\zeta \rangle $ ($U(t)$ being the evolution operator of the
system) remains eigenstate of $b$ in all later time, 
\begin{equation}
b\left\vert \zeta ;t\right\rangle =\zeta (t)\left\vert \zeta ;t\right\rangle
.\text{ }
\end{equation}%
It is clear that the time-evolved states $\left\vert \zeta ;t\right\rangle $
also obey the overcompleteness relation ({\normalsize \ref{compl1}}) and are
eigenstates of the invariant ladder operator $B(t)=U(t)bU^{\dg}(t)$. This
means that the $B(t)$ and $b$ should commute (we suppose that $\zeta (t)$
and $\zeta $ commute). The general form of a fermionic operator is a
(complex) linear combination of $b$, $b^{\dg}$ and $b^{\dg}b$. Such a
combination will commute with $b$ under certain simple restrictions. Taking
into account that the invariants $B(t)$ and $B^{\dg}(t)$ have to obey the
fermion algebra ({\normalsize \ref{013}}) we derive that $[b,B(t)]=0$ if and
only if $B(t)$ is proportional to$b,$ $B(t)=\beta (t)b$. Thus the \textit{%
fermion coherence Hamiltonian} should admit invariant of the form 
\begin{equation}
B_{c}(t)=\beta (t)b,  \label{018}
\end{equation}%
where $\beta (t)$ may be arbitrary complex function of time. As we have
already noted at the end of the preceding section, similar form of the
ladder operator invariant $A_{c}$, eq. ({\normalsize \ref{010}}), is
required in the case of boson systems {\normalsize \cite{Trif75}}. To obtain
now the general fermion coherence Hamiltonian $H_{fCS}$ we apply the
defining requirement for quantum time-dependent invariants $B(t)$, 
\begin{equation}
\frac{\partial }{\partial t}B(t)-i[B(t),H]=0  \label{19}
\end{equation}%
to the operator ({\normalsize \ref{018}}). The general form of fermionic
(one-mode) Hamiltonian is a Hermitian linear combination of $b$, $b^{\dg}$ and 
$b^{\dg}b,$ 
\begin{equation}
H_{\!f}=\omega (t)b^{\dg}b+f(t)b^{\dg}+f^{\ast }(t)b+g(t),  \label{Hf}
\end{equation}%
where $\omega (t)$ and $g(t)$ are real functions of time. The substitution
of this $H_{f}$ into ({\normalsize \ref{19}}) for $B_{c}(t)$ produces the two conditions 
\begin{equation}
\dot{\beta}=i\beta \omega ,\quad 0=\beta f.
\end{equation}%
These simple conditions are readily solved, $f(t)=0$, $\beta (t)=\exp \left(
i\tint_{0}^{t}\omega (\tau )d\tau \right) ,$ leading to Hamiltonian 
\begin{equation}
H_{\!fCS}=\omega (t)b^{\dg}b+g(t),  \label{022}
\end{equation}%
which is the most general form of \textit{fermion coherence Hamiltonian}. If
the evolution of an initial CS $\left\vert \zeta \right\rangle $ is governed
by $H_{fCS}$, then the time-evolved state $\left\vert \zeta ;t\right\rangle $
remains eigenstate of $b$ with eigenvalue 
\begin{equation}
\zeta (t)=\beta ^{-1}(t)\zeta =e^{-i\tint_{0}^{t}\omega (\tau )d\tau }\zeta .
\label{023}
\end{equation}%
The results ({\normalsize \ref{022}}) and ({\normalsize \ref{023}}) are
similar in form, but not identical, to those for the boson systems (%
{\normalsize \ref{Hcs}}) and ({\normalsize \ref{8}}). The fermion coherence
Hamiltonian ({\normalsize \ref{022}}) is of the form of an oscillator with
time dependent frequency (\textit{nonstationary fermion oscillator}), while
the boson coherence Hamiltonian ({\normalsize \ref{Hcs}}) is of the more
general form of the nonstationary \textit{forced} oscillator. In the next
section we find the exact evolution of fermion CS and fermion number states,
governed by the nonstationary forced oscillator Hamiltonian using the
time-dependent integrals of motion method {\normalsize \cite{MMT70, Holz70,
Lewis69}}.

\section{FFO and invariant ladder operators}

We consider the single nonstationary fermionic forced oscillator (FFO)
described by the Hamiltonian ({\normalsize \ref{Hf}}). As we have noted this
in fact is the most general Hamiltonian of single fermion system. The
fermion number operator is defined as $N=b^{\dg}b$. It obey the relation $%
N^{2}=N$ and the three operators $b$, $b^{\dg}$and $N$ close under commutation
the algebra 
\begin{equation}
\left[ b,N\right] =b,\text{ }\left[ b^{\dg},N\right] =-b^{\dg},\text{ }\left[
b,b^{\dg}\right] =1-2N,\text{ \ \ \ }
\end{equation}
The Hilbert space $\mathcal{H}$ of the single-fermion system is spanned by
the two eigenstates $\left\{ \left\vert 0\right\rangle ,\left\vert
1\right\rangle \right\} $ of $N$: 
\begin{equation*}
b^{\dg}b\left\vert n\right\rangle =n\left\vert n\right\rangle ,\text{ \ }n=0,1
\end{equation*}
The operators $b$ and $b^{\dg}$\ allow transitions between the states as 
\begin{equation}
b\left\vert 0\right\rangle =0,\text{\ }b\left\vert 1\right\rangle
=\left\vert 0\right\rangle \,,\text{ \ }b^{\dg}\left\vert 1\right\rangle =0,%
\text{ }b^{\dg}|0\rangle =|1\rangle .\,
\end{equation}
Let us also note that linear combinations of $b^{\dg}$, $b$ and $N$ produce
the half-spin operators $J_{i}$, 
\begin{equation}  \label{Ji}
J_{1}=\tfrac{1}{2}(b^{\dg}+b),\text{ }J_{2}=\tfrac{1}{2i}(b^{\dg}-b),\text{ }
J_{3}=b^{\dg}b-\tfrac{1}{2},  \notag
\end{equation}
closing the \textit{su}(2) algebra: $\left[ J_{k},J_{l}\right] =i\epsilon
_{klm}J_{m}$. It is clear that the fermion forced oscillator Hamitonian $%
H_{fCS}$ belongs to the central extension of \textit{su}(2) (is a linear
combination of $J_{i}$ plus free $C$-number term).

It is convenient to use raising and lowering operators $J_{\pm }=J_{1}\pm
iJ_{2}$ which satisfy the following commutation relation: $\left[ J_{+},
J_{-}\right] =2J_{3},\text{ }\left[ J_{3},J_{\pm }\right] = \pm J_{\pm }%
\text{\ }$, where $J_{+}=b^{\dg},$ $J_{-}=b.$ So that in terms of these half
spin operators the Hamiltonian ({\normalsize \ref{Hf}}) takes the form 
\begin{equation}  \label{Hf 2}
H_{\!f}(t) = \omega (t)J_{3} + f(t)J_{+} + f^{\ast }(t)J_{-} + g(t) + \tfrac{%
\omega (t)}{2}.
\end{equation}

\medskip

Our task is the construction of the time-dependent invariants for the system
(\ref{Hf}), (\ref{Hf 2}). The defining equation of the invariant operator $%
B(t)$ for a quantum system with Hamiltonian $H(t)$ is (\ref{19}). 
Formal solutions to this equation are operators $B(t)=U(t)B(0)U^{\dg}(t)$,
where $U(t)$ is the evolution operator of the system, $U(t)=T\exp
[-i\tint_{0}^{t}H(t^{^{\prime }})dt^{^{\prime }}]$. In our case of FFO (\ref%
{Hf}), (\ref{Hf 2}) we look for the non-Hermitian invariants of the form of
linear combination of the $SU(2)$ generators (\ref{Ji}), 
\begin{equation} \label{B(t)}
\begin{aligned} B(t) = \nu_{-}(t)J_{-} + \nu_{+}(t) J_{+} + \nu_3 (t) J_{3}
,\\ B^\dg(t) = \nu_{-}^* (t)J_{+} + \nu_{+}^*(t) J_{-} + \nu_3^* (t) J_{3},
\end{aligned}  
\end{equation}%
where $\nu _{\pm }(t),\,\,\nu _{3}(t)$ may be complex functions of the time.
Hermitian invariants then can be easily built up as Hermitian combinations
of $B$ and $B^{\dagger}$. In particular if $B$ is a non-Hermitian invariant
the operator 
\begin{equation}
I=B^{\dagger}B-\tfrac{1}{2}  \label{I}
\end{equation}%
is an Hermitian invariant, the fermion analog of the Lewis-Riesenfeld
quadratic invariant \cite{Lewis69}.

Substituting {\normalsize (\ref{B(t)}), (\ref{Ji}) } into {\normalsize (\ref%
{19})}, we find the following system of differential equations for the
parameter functions of $B(t)$ 
\begin{eqnarray}
\dot{\nu}_3 &=& 2i ( \nu _ + f^* - \nu_- f),  \label{(a)} \\
\dot{\nu}_{+} &=& i( \nu_3 f - \nu_+ \omega ),  \label{(b)} \\
\dot{\nu}_{-} &=& i( \nu_- \omega - \nu_3 f^{\ast }).  \label{(c)}
\end{eqnarray}
The solutions of the above linear system of first order equations are
uniquely determined by the initial conditions $\nu_\pm(0)=\nu_{0,\pm}$, $%
\nu_3(0)=\nu_{0,3}$. If we want the invariants $B(t)$ and $B^\dg(t)$ be
again a fermion ladder operator, i.e. to obey the conditions 
\begin{equation}  \label{constr}
B^2(t)= 0,\quad \{B(t),B^\dg(t)\} = 1,
\end{equation}
we have to take $\nu_{0,\pm}$ and $\nu_{0,3}$ satisfying 
\begin{equation}  \label{0}
\nu_{0,3}^2 = -4\nu_{0,+}\nu_{0,-},\quad |\nu_{0,-}| + |\nu_{0,+}| = 1 \, .
\end{equation}
Indeed, for $B^2(t)$ and $\{B(t),B^\dg(t)\}$ we find 
\begin{equation}  \label{lam_i}
\begin{tabular}{l}
$\displaystyle B^2(t) = \nu_+\nu_- +\tfrac{1}{4}\nu_3^2 \equiv
\lambda_1(\nu_\pm,\nu_3)$, \\[2mm] 
$\displaystyle \{B(t),B^\dg(t)\} = |\nu_-|^2 + |\nu_+|^2 + \tfrac 12
|\nu_3|^2 \equiv \lambda_2(\nu_\pm,\nu_3)$.%
\end{tabular}%
\end{equation}
The quantities $\lambda_1(\nu_\pm,\nu_3)$, $\lambda_2(\nu_\pm,\nu_3)$ turned
out to be two different '\textit{constants of motion}' for the system (\ref%
{(a)})-(\ref{(c)}), their time derivatives being vanishing: 
\begin{equation}  \label{dotlam_i}
\begin{tabular}{l}
$\frac{d}{dt}\lambda_1\equiv \frac{d}{dt}\left(\nu_+\nu_- +\tfrac{1}{4}%
\nu_3^2 \right) = 0$, \\[2mm] 
$\frac{d}{dt}\lambda_2\equiv \frac{d}{dt}\left(|\nu_-|^2 + |\nu_+|^2 +
\tfrac 12 |\nu_3|^2 \right) = 0.$%
\end{tabular}%
\end{equation}
Therefore we can fix the values of these constants as $\lambda_1 =0$, $%
\lambda_2=1$, i.e. 
\begin{equation}  \label{lam 0}
\nu_+\nu_- + \tfrac{1}{4}\nu_3^2 = 0,\quad |\nu_-|^2 + |\nu_+|^2 + \tfrac 12
|\nu_3|^2=1,
\end{equation}
and satisfy the conditions (\ref{constr}). If the initial conditions are
taken as 
\begin{equation}  \label{nu_i0}
\nu_-(0) = 1, \,\,\, \nu_+(0) = 0 = \nu_3(0),
\end{equation}
then $B(0)=b$.

Let us first note that in the particular case of the \textit{free oscillator}%
, $f(t)\equiv 0$, the solution of the above system of equations is readily
obtained in the form 
\begin{equation}
\nu _{\pm }(t)=\nu _{0,\pm }e^{\pm i\tint^{t}\omega (\tau )d\tau },\quad \nu
_{3}=\nu _{0,3},  \label{solutions1}
\end{equation}%
where $\nu _{0,\pm }$, $\nu _{0,3}$ are constants. For this solution the
expressions $\lambda _{1}(\nu _{\pm },\nu _{3})$, $\lambda _{2}(\nu _{\pm
},\nu _{3})$ are readily seen to be constant in time as expected: $\lambda
_{1}=\nu _{0,-}\nu _{0,+}+\nu _{0,3}^{2}/4$, $\lambda _{2}=|\nu
_{0,-}|^{2}+|\nu _{0,+}|^{2}+|\nu _{0,3}|^{2}/2$. Then from (\ref{lam_i}), (%
\ref{lam 0}) we see that the invariant fermion annihilation operator $B(t)$
now takes the form $B_{\mathrm{so}}(t)$, 
\begin{equation}
B_{\mathrm{so}}(t)=\nu _{0,-}e^{-i\varphi (t)}b+\nu _{0,+}e^{i\varphi
(t)}b^{\dagger}+2\sqrt{-\nu _{0,-}\nu _{0,+}}\left( b^{\dagger}b-\tfrac{1}{2}%
\right) ,  \label{Bso}
\end{equation}%
where $\varphi (t)=\tint^{t}\omega (\tau )d\tau $ and $|\nu _{0,-}|+|\nu
_{0,+}|=1$, the phases of $\nu _{0,\pm }$ remaining arbitrary.

Consider now in greater detail the nonstationary \textit{forced oscillator}
with nonvanishing $f(t)$: $f(t)\neq 0$.

For this system we can express all the three parameter functions $\nu _{\pm
}(t)$, $\nu _{3}(t)$ in terms of one of them, which has to obey more simple
second order differential equation. Let for example, express $\nu _{3}(t)$
and $\nu _{-}(t)$ in terms of $\nu _{+}(t)$ and its derivatives. We have 
\begin{equation}
\nu _{3}=-\tfrac{i}{f}(\dot{\nu}_{+}+i\nu _{+}\omega ),  \label{nu3}
\end{equation}%
\begin{equation}
\nu _{-}=\tfrac{1}{2f^{2}}\left[ \ddot{\nu}_{+}+\left( i\omega -\tfrac{\dot{f%
}}{f}\right) \dot{\nu}_{+}+\left( 2ff^{\ast }+i\dot{\omega}-i\frac{\omega }{f%
}\dot{f}\right) \nu _{+}\right] .  \label{nu-}
\end{equation}%
Taking the time derivative of both sides of (\ref{nu-}) and using eqs. (\ref%
{(c)}) and (\ref{nu3}) we arrive to the third order equation for $\nu
_{+}(t) $, 
\begin{eqnarray}\label{nu+}
\tfrac{1}{2f^{2}}\dddot{\nu}_{+} &= \tfrac{3\dot{f}}{2f^{3}}\ddot{\nu}%
_{+}-\left( 2\tfrac{f^{\ast }}{f}+\tfrac{\omega ^{2}}{2f^{2}}+\tfrac{i\dot{%
\omega}}{2f^{2}}+\tfrac{3\dot{f}^{2}}{2f^{4}}-i\tfrac{\omega \dot{f}}{f^{3}}-%
\tfrac{\ddot{f}}{2f^{3}}\right) \dot{\nu}_{+} -  \notag   \\
& \left( \tfrac{\dot{f}^{\ast}}{f} + \tfrac{\omega\dot{\omega}}{2f^{2}%
} - \tfrac{\omega^{2}\dot{f}}{2f^{3}} - \tfrac{f^{\ast }\dot{f}}{f^{2}} - i\tfrac{%
3\dot{\omega}\dot{f}}{2f^{3}} + \tfrac{i\ddot{\omega}}{2f^{2}} - \tfrac{i\omega\ddot{f}}{2f^{3}} + i\tfrac{3\omega \dot{f}^{2}%
}{2f^{4}}\right) \nu _{+}\,.
\end{eqnarray}%
Furthermore if we could find \textit{a first integral} of the equation (\ref%
{nu+}) then we can express $\nu _{-}$ in terms of $\nu _{+},\,\,\dot{\nu}%
_{+} $, eliminating the second derivative in eq. (\ref{nu-}). One can check
that the following expression of $\nu _{+},\,\dot{\nu}_{+}$ and $\ddot{\nu}%
_{+}$ is a first integral of equation (\ref{nu+}) (that is $d\lambda /dt=0$%
), 
\begin{equation}
\lambda =\tfrac{4}{f^{2}}\left[ 2\nu _{+}\ddot{\nu}_{+}-\dot{\nu}%
_{+}^{2}-2\nu _{+}\dot{\nu}_{+}\tfrac{\dot{f}}{f}+4\nu _{+}^{2}\left(
|f|^{2}+\tfrac{\omega ^{2}}{4}+i\tfrac{\dot{\omega}}{2}-i\tfrac{\omega \dot{f%
}}{2f}\right) \right] .  \label{lam}
\end{equation}%
We regard this formula as second order equation for $\nu _{+}$, depending on
an arbitrary constant $\lambda $. Using this, and supposing that $\nu
_{+}\neq 0$, we obtain for $\nu _{-}$ the more compact expression in terms
of $\nu _{+}$ and $\dot{\nu}_{+}$, 
\begin{equation}
\nu _{-}=\tfrac{\lambda }{16\nu _{+}}-\tfrac{1}{4f^{2}\nu _{+}}\left( \omega
\nu _{+}-i\dot{\nu}_{+}\right) ^{2},  \label{nu- 2}
\end{equation}%
and we see that $\nu _{-}=(\lambda /4-\nu _{3}^{2})/4\nu _{+}$. The first
integral $\lambda $ of eq. (\ref{nu+}) is proportional to the constant of
motion $\lambda _{1}$ of system (\ref{(a)}) - (\ref{(c)}): $\lambda
_{1}=\lambda /16$.

Thus the operators $B(t),\,B^{\dagger}(t)$, eq.(\ref{B(t)}), are invariant
for the forced oscillator (\ref{Hf}), (\ref{Hf 2}) if $\nu _{+}(t)$ is a
nonvanishing solution of the second order equation (\ref{lam}) with any
constant $\lambda $, and $\nu _{3}$ and $\nu _{-}$ being given by eqs. (\ref%
{nu3}) and (\ref{nu- 2}) respectively. They will obey the fermionic ladder
operator conditions (\ref{constr}) if $\lambda _{1}=0=\lambda $ and $\lambda
_{2}=1$. Instead of fixing $\lambda _{2}=1$ we can redefine $B(t)\rightarrow
B(t)/\sqrt{\lambda_{2}}$, i.e. take the invariant fermion annihilation operator of
the form valid for any nonnegative constant of motion $\lambda _{2}=|\nu
_{-}|^{2}+|\nu |^{2}+|\nu _{3}|^{2}/2$, 
\begin{equation}
B(t)=\tfrac{1}{\sqrt{\lambda _{2}}}\left[ \tfrac{1}{f}\left( \nu _{+}\omega
\!-\!i\dot{\nu}_{+}\right) (b^{\dagger}b-\tfrac{1}{2})+\nu _{+}b^{\dagger}-%
\tfrac{1}{4f^{2}\nu _{+}}\left( \omega \nu _{+}\!-\!i\dot{\nu}_{+}\right)
^{2}b\right] \, ,  \label{B(t) 2}
\end{equation}%
where $\nu _{+}$ is a solution to the equation (\ref{lam}) with $\lambda =0$.

In this case the equations (\ref{nu+}), (\ref{nu- 2}) and (\ref{nu3}) can be
greatly simplified if we put 
\begin{equation}
\nu _{+}(t)=\tfrac{1}{2}\epsilon ^{2}(t).  \label{nu+ eps}
\end{equation}%
The result is 
\begin{equation}
\begin{tabular}{l}
$\displaystyle\nu _{-}=-\frac{1}{2f^{2}}\left( \tfrac{\omega }{2}\epsilon -i%
\dot{\epsilon}\right) ^{2}$, \\ 
$\displaystyle\nu _{3}=\frac{1}{f}\left( \tfrac{\omega }{2}\epsilon
^{2}-i\epsilon \dot{\epsilon}\right) $,%
\end{tabular}%
\end{equation}%
where $\epsilon (t)$ satisfies the equation 
\begin{equation}
\ddot{\epsilon}-\tfrac{\dot{f}}{f}\dot{\epsilon}+\Omega (t)\epsilon =0,
\label{eps eq}
\end{equation}%
\begin{equation}
\Omega (t)=|f(t)|^{2}+\tfrac{1}{4}\omega ^{2}(t)+\tfrac{i}{2}\dot{\omega}-%
\tfrac{i}{2}\omega \tfrac{\dot{f}}{f}.
\end{equation}%
This latter equation admits $\lambda _{2}$, eq. (\ref{lam_i}), as a first
integral, 
\begin{equation}
\lambda _{2}=\tfrac{|\epsilon |^{4}}{4}\left( 1+\tfrac{2}{|f|^{2}}\left\vert 
\tfrac{\omega }{2}\epsilon -i\dot{\epsilon}\right\vert ^{2}+\tfrac{1}{|f|^{4}%
}\left\vert \tfrac{\omega }{2}\epsilon -i\dot{\epsilon}\right\vert
^{4}\right) .
\end{equation}%
In (\ref{eps eq}) the term proportional to the first derivative can be
eliminated by the substitution 
\begin{equation}
\epsilon =\epsilon ^{\prime }\exp \left( \tfrac{1}{2}\tint^{t}\dot{f}(\tau
)d\tau /f(\tau )\right) ,
\end{equation}%
which leads to the more simple equation 
\begin{equation}
\ddot{\epsilon}\,^{\prime }+\Omega ^{\prime }(t)\epsilon ^{\prime }=0.
\label{eps'}
\end{equation}%
\begin{equation}
\Omega ^{\prime }(t)=|f(t)|^{2}+\tfrac{1}{4}\omega ^{2}(t)+\tfrac{i}{2}\dot{%
\omega}-\tfrac{i}{2}\omega \tfrac{\dot{f}}{f}+\tfrac{1}{2}(\frac{\ddot{f}}{f}%
)-\tfrac{3}{4}(\frac{\dot{f}\text{ }^{2}}{f^{2}}).
\end{equation}%
It is worth noting at this point that the invariant ladder operators for the
boson nonstationary forced oscillator have been obtained \cite{MMT70, Holz70}
in terms of solutions to the same classical equation (\ref{eps'}).

\section{CS for the fermion forced oscillator}

We define coherent states (CS) for a given fermion system as eigenstates of
the corresponding invariant fermion annihilation (or creation) operator $%
B(t) $. Since the most general fermion one mode Hamiltonian operator is of
the form of (nonstationary) forced oscillator (\ref{Hf 2}), the one-mode
fermion CS are defined as eigenstates of the invariant ladder operator $B(t)$%
, eq. (\ref{B(t) 2}) (or eqs. (\ref{B(t)}), (\ref{constr})): 
\begin{equation}
B(t)|\zeta ;t\rangle =\zeta |\zeta ;t\rangle .  \label{|z;t> 1}
\end{equation}%
Since $B(t)$ is invariant operator for the FFO, the eigenvalue $\zeta $ does
not depend on time $t$. In terms of the $\zeta $, $B(t)$, $B^{\dagger }(t)$
and the $B(t)$-vacuum $|0;t\rangle $ we have for $|\zeta ;t\rangle $ the
same formulas as for the canonical fermion CS $|\zeta \rangle $, eq. (\ref%
{b|z>}), (\ref{|z>})-(\ref{z ints}), in terms of $\zeta $, $b$, $b^{\dagger
} $ and the $b$-vacuum $|0\rangle $. In particular 
\begin{equation}
|\zeta ;t\rangle =e^{-\tfrac{1}{2}\zeta ^{\ast }\zeta }\left( |0;t\rangle
-\zeta B^{\dagger }(t)|0;t\rangle \right) .  \label{|z;t> 2}
\end{equation}%
It remains therefore to construct the (normalized) new ground state $%
|0;t\rangle $ according to its defining equations 
\begin{equation}
\begin{aligned} B(t) |0;t\rangle = 0 ,\\ i\frac{d}{dt} |0;t\rangle =
H_{f}|0;t\rangle. \end{aligned}  \label{|0;t> 1}
\end{equation}%
We put 
\begin{equation}
|0;t\rangle =\alpha _{0}(t)|0\rangle +\alpha _{1}(t)|1\rangle ,
\label{|0;t> 2}
\end{equation}%
substitute this into (\ref{|0;t> 1}) and after some tedious calculations
find 
\begin{equation}
\begin{aligned} \alpha_1(t) &= \alpha_0(t)\frac{\nu_3^*(t)}{2\nu_+^*(t)},\\
\alpha_0(t) &=
\sqrt{|\nu_+(t)|}\exp\left[-\tfrac{i}{2}\left(\varphi_{\nu_+}(t) +
\tint^t\left(2g(\tau) + \omega(\tau)\right)d\tau\right)\right], \end{aligned}
\label{alf_i}
\end{equation}%
where $\varphi _{\nu _{+}}$ is the phase of $\nu _{+}(t)$. The state $|\zeta
;t\rangle $ will represent the exact time evolution of an initial canonical
CS $|\zeta \rangle $ if the initial conditions (\ref{nu_i0}) are imposed: $%
|\zeta ;0\rangle =|\zeta \rangle $. In this case, as we have shown in
section 3, the time evolved state $|\zeta ;t\rangle $ could be again an
eigenstate of $b$ if the oscillator is not 'forced', i.e. if $f(t)=0$. Let
us note that the time-dependence of the constructed states is obtained in
terms of solutions to the classical system (\ref{(a)})-(\ref{(c)}), or
equivalently to the classical equation (\ref{eps'}). The latter is the same
equation that appeared in the time evolution of the CS of bosonic FFO \cite%
{MMT70, Holz70}

Our method of construction of dynamical invariants differs slightly from the
Lewis-Riesenfeld method \cite{Lewis69} (developed for bosonic oscillators).
Lewis and Riesenfeld used to first construct Hermitian invariant, which then
is represented as a product of normally ordered ladder operators. To make
connection to their approach let us suppose that we first succeeded to
construct the Hermitian invariant $N(t)$ and to find some ladder operators $%
\tilde{B}(t)$, $\tilde{B}^{\dagger}(t)$ that factorize it: $N(t)=\tilde{B}%
^{\dagger}(t)\tilde{B}(t)$. It is clear that $\tilde{B}(t)$ may differ from
our non-Hermitian invariant $B(t)$ in a phase factor:\thinspace\ $\tilde{B}%
(t)=e^{i\varphi (t)}B(t)$.\newline
We can then in a standard way construct normalized eigenstates of $N(t)$, 
\begin{equation}
N(t)\widetilde{|0;t\rangle }=0\quad \mathrm{and}\quad N(t)\widetilde{%
|1;t\rangle }=\widetilde{|1;t\rangle },  \label{tld 2}
\end{equation}%
and of $\tilde{B}(t)$, \ 
\begin{equation}
\tilde{B}(t)\widetilde{|\zeta ;t\rangle }=\zeta \widetilde{|\zeta ;t\rangle }%
,
\end{equation}
\begin{equation}
\widetilde{|\zeta ;t\rangle }=\left( 1-\tfrac{1}{2}\zeta ^{\ast }\zeta
\right) \left[ \widetilde{|0;t\rangle }-\zeta \widetilde{|1;t\rangle }\right]
\label{tld 3}
\end{equation}%
which however do not obey the Schr\"{o}dinger equation since, in general $%
\tilde{B}(t)$ may not be invariant. To obtain solutions $|n;t\rangle $ and $%
|\zeta ;t\rangle $ the above eigenstates $\widetilde{|n;t\rangle }$, $n=0,1$%
, should be multiplied by phase factors, 
\begin{equation}
|n;t\rangle =e^{i\phi _{n}(t)}\widetilde{|n;t\rangle },\quad n=0,1,
\label{tld 4}
\end{equation}%
\begin{equation}
|\zeta ;t\rangle =\left( 1-\tfrac{1}{2}\zeta ^{\ast }\zeta \right) \left[
e^{i\phi _{0}(t)}\widetilde{|0;t\rangle }-\zeta e^{i\phi _{1}(t)}\widetilde{%
|1;t\rangle }\right]  \label{tld 5}
\end{equation}%
which should obey the equations \ 
\begin{equation}
\tfrac{d}{dt}\phi _{n}=\widetilde{\langle n;t|}i\tfrac{\partial }{\partial t}%
-H\widetilde{|n;t\rangle }.  \label{tld 6}
\end{equation}%
Evidently the state (\ref{tld 5}) is eigenstate of $\tilde{B}(t)$ with time
dependent eigenvalue $\zeta (t)=\exp (i\varphi (t))$, $\varphi (t)=\phi
_{1}(t)-\phi _{0}(t)$.\newline
The phase $\varphi (t)=\phi _{1}(t)-\phi _{0}(t)$ consists of two parts -
geometrical one $\varphi ^{G}$, and dynamical one $\varphi ^{D}=\varphi
-\varphi ^{G}$ \cite{Maam99}, 
\begin{eqnarray}
\varphi ^{G}(t) &=&i\int_{0}^{t}\left( \widetilde{\langle 1;t^{\prime }|}%
\frac{\partial }{\partial t^{\prime }}\widetilde{|1;t^{\prime }\rangle }-%
\widetilde{\langle 0;t^{\prime }|}\frac{\partial }{\partial t^{\prime }}%
\widetilde{|0;t^{\prime }\rangle }\right) dt^{\prime }  \label{tld 8} \\
&=&\varphi (t)+\int_{0}^{t}\left( \widetilde{\langle 1;t^{\prime }|}H%
\widetilde{|1;t^{\prime }\rangle }-\widetilde{\langle 0;t^{\prime }|}H%
\widetilde{|0;t^{\prime }\rangle }\right) dt^{\prime }
\end{eqnarray}

\section*{Concluding Remarks}

\ \ \ In this article, we have extend the earlier results of the \textit{%
boson} coherence Hamiltonian and boson invariant ladder operators to the 
\textit{fermion} coherence Hamiltonian and fermion invariant ladder
operators.\textit{\ } We have pointed out that unlike the boson coherence
Hamiltonian, which is of the more general form of the nonstationary \textit{%
forced} oscillator, the fermion coherence Hamiltonian is of more simple form
of (nonstationary) fermion oscillator.

For the more general (in fact most general) fermionic system of
nonstationary forced oscillator we have constructed invariant ladder
operators and the related Fock and coherent states. We succeeded to express
these invariants and the time evolution of the corresponding states in terms
of the same classical equation, that describe the evolution of coherent
states of the boson nonstationary forced oscillator \cite{MMT70, Holz70}.
The relation of the invariant ladder operators method to the
Lewis-Riesenfeld method \cite{Lewis69} was briefly described on the example
of nonstationary fermion systems.

\end{document}